\documentclass[aps,nofootinbib,superscriptaddress,twocolumn,floatfix]{revtex4}
\usepackage{graphicx}
\usepackage{amsmath}
\usepackage{dcolumn}
\usepackage{epstopdf}
\usepackage{lipsum}

\begin{document}

\title{Strong Coupling in Hyperbolic Metamaterials}
\author{Prashant Shekhar}
\author{Zubin Jacob} 
\affiliation{Department of Electrical and Computer Engineering, University of Alberta, Edmonton, AB T6G 2V4, Canada}
\begin{abstract}


Nanoscale light-matter interaction in the weak coupling regime has been achieved with unique hyperbolic metamaterial modes possessing a high density of states. Here, we show strong coupling between intersubband transitions (ISBTs) of a multiple quantum well (MQW) slab and the bulk polariton modes of a hyperbolic metamaterial (HMM).  These HMM modes have large wavevectors (high-$k$ modes) and are normally evanescent in conventional materials.  We analyze a metal-dielectric practical multilayer HMM structure consisting of a highly doped semiconductor acting as a metallic layer and an active  multiple quantum well dielectric slab. We observe delocalized metamaterial mode interaction with the active materials distributed throughout the structure. Strong coupling and characteristic anticrossing with a maximum Rabi splitting (RS) energy of up to 52 meV is predicted between the high-$k$ mode of the HMM and the ISBT, a value approximately 10.5 times greater than the ISBT linewidth and 4.5 times greater than the material loss of the structure. The scalability and tunability of the RS energy in an active semiconductor metamaterial device have potential applications in quantum well infrared photodetectors and intersubband light-emitting devices.


\end{abstract}
\maketitle

\section{Introduction}

Metamaterials, artificial media synthesized from nanostructured building blocks, have recently shown promise for engineering nanoscale light-matter interactions \cite{shalaev_optical_2007,menon_negative_2008,foteinopoulou_refraction_2003,jacob_optical_2006,liu_far-field_2007,zayats_plasmonic,govyadinov_metamaterial_2006}. This has opened the possibility for quantum applications with metamaterials \cite{tanaka2010multifold,jacob_plasmonics_2011} and, in particular, modified spontaneous emission of quantum emitters in the weak coupling limit (irreversible regime) \cite{jacob_broadband_2012,noginov_controlling_2010,lu_enhancing_2014,ferrari_enhanced_2014}. The two signatures of the modified spontaneous emission are the reduced lifetime and altered far field emission pattern \cite{ni2011effect}. Strong coupling, unlike the weak coupling limit, relies on the back-action between the emitter and the metamaterial to create coherent states between light and matter \cite{khitrova_vacuum_2006,ameling_strong_2011,rogacheva_giant_2006}. The features of strong coupling are often ascertained through the spectral signatures in either the absorption or emission of the coupled emitter-environment system \cite{andreani1999strong}. Additionally, strong coupling is characterized by the temporal dynamics of the energy oscillations between the emitter and photonic mode of the system \cite{reithmaier_strong_2004}.

Microcavities, nanocavities and photonic crystals have been studied extensively with both weakly coupled \cite{michler_quantum_2000,pelton_efficient_2002,lodahl_controlling_2004,hughes_enhanced_2004} and strongly coupled \cite{khitrova_vacuum_2006,ameling_strong_2011,rogacheva_giant_2006,shelton_strong_2011,steinbach_electron-phonon_1999,reithmaier_strong_2004,feuillet-palma_extremely_2012,scalari_ultrastrong_2013,benz_strong_2013,weis_hybridization_2011,sapienza_photovoltaic_2007} emitters. Although the strongly coupled systems are able to effectively couple light and matter, they are often wavelength sized diffraction limited structures. Additionally, the resonant nature of the modes limits their bandwidth of operation. Propagating surface-plasmon polaritons on metals are a suitable candidate for subwavelength radiative decay engineering \cite{lakowicz2004radiative} or strong coupling \cite{dintinger2005strong,bellessa2004strong,christ2003waveguide,trugler_strong_2008,delteil_charge-induced_2012,wurtz_molecular_2007} while low mode volume surface plasmons (eg: 1D nanowire) can be used for broadband coupling between emitters and plasmons \cite{chang2005quantum}. 

A natural question then arises whether delocalized plasmonic modes which lead to collective metamaterial behavior can show effects such as coherence and strong coupling. Resonant metamaterials have shown strong coupling to quantum well emitters proving that even lossy modes can enter the strong coupling regime \cite{meinzer2010arrays}. Here, we introduce strong coupling at the nanoscale with non-resonant hyperbolic metamaterials (HMMs). HMMs are a special class of metamaterial with an extremely anisotropic dielectric tensor resulting in a hyperbolic dispersion for the structure \cite{smith2003electromagnetic,podolskiy2005strongly,poddubny_hyperbolic_2013,govyadinov_metamaterial_2006,jacob_broadband_2012} and have shown promise in the weak coupling regime \cite{cortes_quantum_2012}, specifically in the field of radiative decay engineering to produce broadband single-photon sources \cite{jacob_broadband_2012,noginov_controlling_2010,Guo2012,jacob_plasmonics_2011,newman2013enhanced,iorsh2012spontaneous}. In this paper, we predict strong coupling behaviour with the subwavelength modes of an HMM. We show that such strong coupling effects can persist even in the presence of metallic losses. We provide a practical semiconductor superlattice design for our metamaterial consisting of highly doped n$^+$-In$_{0.53}$Ga$_{0.47}$As as the metallic building block and an embedded active multiple quantum well layer (Al$_{0.35}$Ga$_{0.65}$As/GaAs). The proposed structure can be fabricated by molecular beam epitaxy grown on lattice matched InP substrates \cite{hoffman_negative_2007} and the predicted effect can be isolated in experiment through angle resolved spectroscopy of the quantum well absorption. Additionally, our proposed structure can show effects with enhanced nonlinearities and polariton interaction due to nanoscale strong coupling.  

This work presents the initial steps to realizing novel mixed and coherent states between metamaterial modes and embedded emitters. In the limit of many quantum emitters in a system (eg: multiple quantum wells, thin film of dye molecules or quantum dots), strong coupling behaviour in metamaterial structures can be treated semiclassically. However, single emitter systems can show anharmonic effects which require a fully quantized treatment \cite{bishop2008nonlinear,zhu1990vacuum,kasprzak2010up}. The same holds true in the weak coupling regime where effects such as antibunching of light from isolated emitters cannot be treated classically \cite{lounis2005single}. Experimental verification of strong coupling in the semiclassical regime between quantum wells and hyperbolic metamaterial states should lead to avenues of realizing quantum strong coupling with single emitters and metamaterials.

\section{Hyperbolic Metamaterials}
 \subsection{Semiconductor HMMs}
 \label{HMMs}
Hyperbolic Metamaterials (HMMs) are artificial uniaxial materials with an extremely anisotropic dielectric tensor. The extreme anisotropy requires the components of the permittivity to be defined such that $\epsilon_{xx}= \epsilon_{yy}$ and $\epsilon_{zz}\times\epsilon_{xx} < 0$. The unique electromagnetic response gives rise to an unconventional dispersion relation for extraodinary waves in a uniaxial material: $\frac{k_x^2+k_y^2}{\epsilon_{zz}}+\frac{k_z^2}{\epsilon_{xx}} =\left(\frac{\omega}{c}\right)^2$. The term hyperbolic is used to describe the hyperbolic dispersion of the isofrequency surface of the HMM as opposed to the spherical or ellipsoidal isofrequency surfaces seen in conventional materials. The HMM can support waves with large wavevectors (high-$k$ waves) as a result of its characteristic hyperbolic dispersion \cite{cortes_quantum_2012, Guo2012,jacob_optical_2006,govyadinov_metamaterial_2006,liu_far-field_2007}.

One realization of a hyperbolic metamaterial involves a planar multilayer structure with alternating subwavelength metal-dielectric layers \cite{jacob_optical_2006,liu_far-field_2007}. The high-$k$ modes of the system arise from the near-field coupling of the surface plasmon polaritons (SPPs), excited with incident p-polarized light, at each of the metal-dielectric interfaces in the structure. The high-$k$ modes are the Bloch modes of the metal-dielectric superlattice.  \cite{jacob_broadband_2012,jacob_plasmonics_2011}. 

Degenerately doped semiconductors have plasmonic resonances in the mid-IR that can replace the metal in a conventional metal-dielectric HMM to create a new class of semiconductor HMMs \cite{hoffman_negative_2007}. These semiconductor HMMs, aside from their ability to support high-$k$ states in the near-IR and mid-IR, have the distinct advantage of being able to tune their plasma frequencies by variation of their electron doping density.

\begin{figure*}[ht]
	\includegraphics{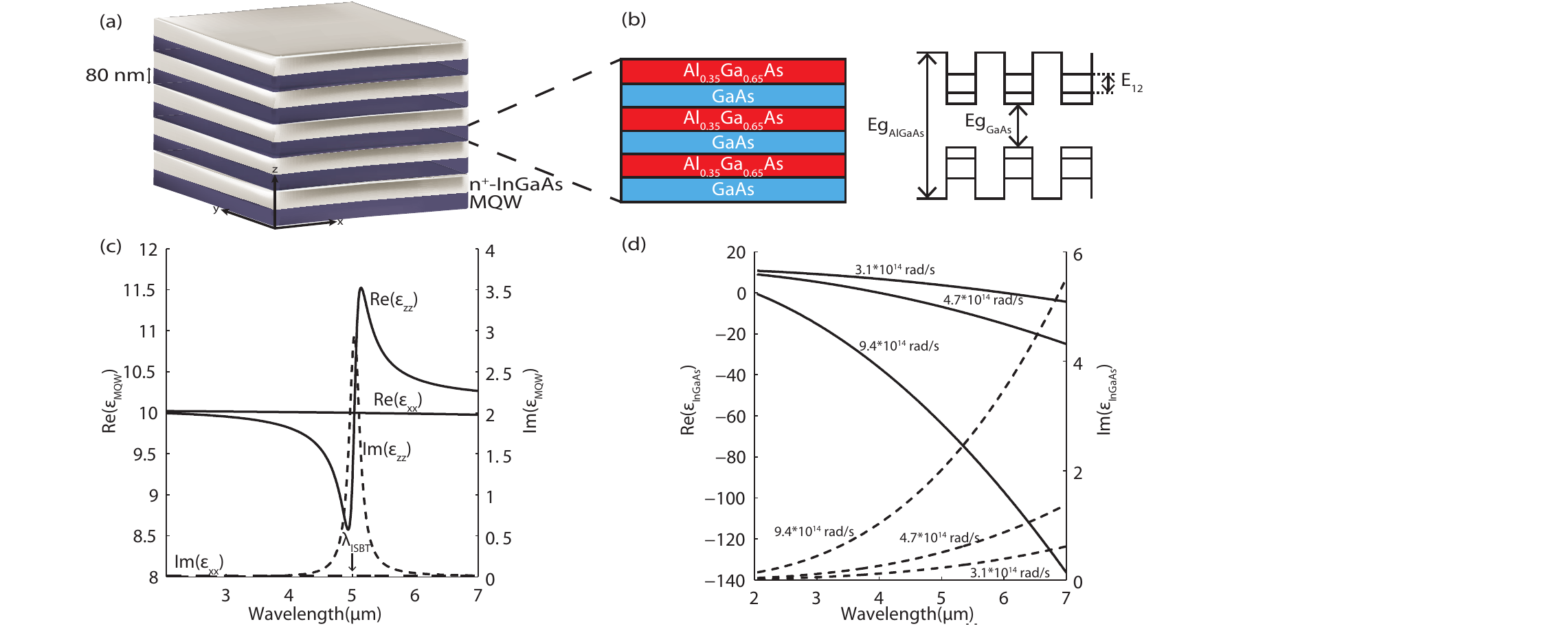}
	\caption{(a) Multilayer realization of a semiconductor HMM. The structure consists of alternating 80 nm thick subwavelength layers of a dielectric MQW slab and degenerately doped n$^+$-In$_{0.53}$Ga$_{0.47}$As. The extreme anisotropy of the structure results in a hyperbolic isofrequency surface. (b) Quantum well structure of the MQW slab. (c) Perpendicular and parallel permittivities of the purely dielectric MQW slab. The resonance at a wavelength of $\lambda_{ISBT}=5 \mu$m corresponds to the energy of the intersubband transition in a single quantum well of the MQW structure. (d) Dispersion of the n$^+$-In$_{0.53}$Ga$_{0.47}$As semiconductor layer at different plasma frequencies. The negative permittivity of the In$_{0.53}$Ga$_{0.47}$As layer (metallic response) is required in order to realize a hyperbolic isofrequency surface for the semiconductor HMM.}
	\label{Slab}
\end{figure*}

The plasmonic semiconductor, for example, can be a degenerately doped In$_{0.53}$Ga$_{0.47}$As (n$^+$-In$_{0.53}$Ga$_{0.47}$As) semiconductor with $Re(\epsilon_{InGaAs})<0$. The n$^+$-In$_{0.53}$Ga$_{0.47}$As layer is assumed to be isotropic and approximated with Drude-like behaviour in the following manner \cite{hoffman_negative_2007}: $\epsilon_{InGaAs} = \epsilon_{b,InGaAs}\left(1-\frac{\omega_{p,InGaAs}^2}{\omega^2+i\omega\gamma}\right)$. Here, $\epsilon_{b,InGaAs}$ is the background dielectric set at 12.15, $\gamma$ is the electron scattering rate set to $1\times10^{13} s^{-1}$ and $\omega_{p,InGaAs}$ is the plasma frequency. Figure \ref{Slab}(d) shows the dispersion of $\epsilon_{InGaAs}$ at different plasma frequencies of the semiconductor. For the analysis in this paper, we set $\omega_{p,InGaAs}=9.43\times10^{14}rad/s$ (corresponding to a doping density of $2.5\times10^{19}cm^{-3}$) to best interact with the dielectric component of our structure in the mid-IR. The proposed design can also be tuned to be effective at longer wavelengths in the mid-IR where degenerately doped semiconductors can be easily achieved due to lower plasma frequencies and thus reduced doping density requirements.

\subsection{Dielectric Component: Effective Medium Theory for the Multiple Quantum Well Slab}
\label{EMTMQW}

In this paper, we look at a semiconductor HMM with a multiple quantum well (MQW) slab acting as an active dielectric layer. The MQW slab itself is modeled with an effective medium theory (EMT) approach. Previous analytic and experimental work have shown the validity of using EMT to model the behaviour of MQW structures \cite{agranovich_cavity_2003,plum}. The quantum well thickness, $L_{QW}$ (6 nm), and the MQW period, $L_{MQW}$ (20 nm), are much smaller than the wavelength of the incident infrared radiation and the wells themselves are assumed to be quantum mechanically isolated from each other. This is the first scale of homogenization for the 80 nm thick MQW structure. We will see, in Section \ref{SCEMT}, that a second homogenization between the MQW and the metallic components will be performed to describe the overall metal-dielectric effective medium. The MQW slab is composed of alternating subwavelength layers of Al$_{0.35}$Ga$_{0.65}$As and GaAs to form the multiple quantum wells. Here, Al$_{0.35}$Ga$_{0.65}$As, with its larger bandgap relative to GaAs, creates the barriers for the structure (Figure \ref{Slab}(b)).

MQWs show free electron movement in the plane parallel to the surface (the x-y plane) and quantum confinement, with possible intersubband transitions (ISBTs), in the the plane normal to the interface (z-direction). This quasi-two-dimensional electron gas can be modeled with an anisotropic dielectric tensor with a uniaxial crystal symmetry. The permittivity of the MQW slab, in the plane parallel to the interface, is effectively characterized by a Drude model \cite{plum,zaluzny_coupling_1999}: 

\begin{equation}
	\epsilon_{xx}^d =\epsilon_{yy}^d=\epsilon_y-\frac{\omega_{p,mqw}^2}{\omega^2 +i\omega\gamma_1}
	\label{Drude}
\end{equation}

Permittivity in the plane perpendicular to the MQW interface is characterized with a Lorentzian Oscillator Model in order to incorporate the quantum confinement effects of the structure, specifically the resonance at the ISBT energy \cite{plum,zaluzny_coupling_1999}:

\begin{equation}
	\frac{1}{\epsilon_{zz}^d}=\frac{1}{\epsilon_z} -\frac{\frac{\omega_{p,mqw}^2f_{12}}{2\omega\gamma_2\epsilon_{well}}}{\frac{E_{12}^2-\hbar^2\omega^2}{2\hbar^2\gamma_2\omega}-i}
	\label{LOModel}
\end{equation}

Here, $\epsilon_y$ and $\epsilon_z$ represent the mean effective background dielectric constant and are given as $\epsilon_y = (1-L_{QW}/L_{MQW})\epsilon_{barrier}$ and $\epsilon_z^{-1} = (1-L_{QW}/L_{MQW})/ \epsilon_{barrier} + (L_{QW}/L_{MQW})/\epsilon_{well}$, with $\epsilon_{barrier}$ (9.88) and $\epsilon_{well}$(10.36) representing the undoped background dielectric constant for the barrier and well respectively \cite{zaluzny_coupling_1999}. $\omega_{p,mqw} = (n_se^2/m\epsilon_0L_{MQW})^{1/2}$ is the plasma frequency for the system where $e$ is the elementary charge of the electron, $\epsilon_0$ is the vacuum permittivity constant, m* = 0.0665m is the effective mass of the electron, where m is the mass of an electron in vacuum, and $n_s = 1.5\times10^{12} cm^{-2}$ is the areal electron density per quantum well. $f_{12}$ corresponds to the oscillator strength of the resonance which depends on the transition energy of the ISBT, the effective mass of the electron and the intersubband dipole matrix element. $E_{12}$ is the ISBT transition energy, which, for the parameters established is set to be equal to $\lambda_{ISBT}=5\mu$m (0.2480 eV). $\gamma_1$ = $\gamma_2$ is the electron scattering rate given as $7.596\times10^{12}s^{-1}$ \cite{plum}.

The superscript $d$ in the definitions of both the parallel and perpendicular permitivitties given in Equation \ref{Drude} and Equation \ref{LOModel} respectively, is used to emphasize that the MQW slab is purely dielectric as we are operating above the plasma frequency ($\omega_{p,mqw}$) of the MQW slab. The dispersion for the MQW slab can be seen in Figure \ref{Slab}(c).

One will note that the properties of the ISBT are only present in the perpendicular ($\epsilon_{zz}$) component of the dielectric tensor (Equation \ref{LOModel}) as the ISBT can only be excited with electric fields polarized in the growth direction (z-direction) of the slab. This is due to the fact that wavefunctions of each subband are also bound in the z-direction and thus, due to orthogonality conditions, transitions between states require absorption from z-polarized E-fields. S-polarized light will have no z-component of the electric field, and thus p-polarized incidence is required to see effects of the ISBTs in the semiconductor HMM.

With p-polarized plane wave incidence, this semiconductor HMM can support both ISBTs and high-$k$ modes simultaneously. In this paper, we show that the high-$k$ waves of the semiconductor HMM strong couple to intersubband transitions (ISBTs) present in the dielectric MQW layer.

\subsection{Basis of Strong Coupling: Semi-classical Perspective}
\label{BasisSC}

Strong coupling has been a key area of interest over the past decade for its potential in creating coherent and entangled states between light and matter \cite{khitrova_vacuum_2006,ameling_strong_2011,rogacheva_giant_2006,shelton_strong_2011,steinbach_electron-phonon_1999}. It is the result of a large interaction between two distinct resonances within a system. In a semiconductor HMM, for example, strong light-matter coupling is possible between the Lorentzian resonance of the ISBTs and the high-$k$ modes of the structure. Strong coupling between two resonances results in a typical polaritonic dispersion and a collective excitation unlike the weak coupling limit \cite{nov}. Specifically, in the strong coupling regime, the strength of coupling between the two resonances is greater than the sum of the damping rates of both resonators. We will first derive semiclassical strong coupling behaviour between a Lorentzian resonance, such as an ISBT, and a ‘high-$k$’ mode of an HMM. 

\par Using the dispersion relation for extraodinary waves in a uniaxial medium given at the beginning of Section \ref{HMMs}, we define the energy for a high-$k$ mode in a semiconductor HMM as:

\begin{equation}
	E_{high-k}^2(q) = \hbar^2c^2\left(\frac{q^2}{\epsilon_{zz}}+\frac{k_z^2}{\epsilon_{xx}}\right)
	\label{high-kEnergy}
\end{equation}

Here $q^2 = k_x^2+ k_y^2$ , $\epsilon_{zz}$ and $\epsilon_{xx}$ are the perpendicular and parallel permittivity respectively, and $k_z$ is the wavevector normal to the interface. We now assume a Lorentzian ISBT resonance is added to the HMM in the form of a low loss Lorentzian Oscillator model. The dispersion given in Equation \ref{high-kEnergy} can now be rewritten as follows \cite{agranovich_cavity_2003}: 

\begin{equation}
	\left(\frac{\hbar^2c^2q^2}{E^2-\frac{\hbar^2c^2k_z^2}{\epsilon_{xx}}}\right)=\epsilon_{zz}+\frac{C}{E_{ISBT}^2-E^2}
	\label{AddingL0}
\end{equation}

Note that we are only adding the ISBT resonance to the perpendicular ($\epsilon_{zz}$) component of the permittivity since the absorption requires the field component perpendicular to the growth axis (Section \ref{EMTMQW}). C is the constant representative of the oscillator strength of the resonance and $E_{ISBT}$ is the ISBT energy. In the regime of strong coupling, we assume that the resonant energy ($E_{ISBT}$) and the high-$k$ mode energy ($E_{high-k}$) become degenerate such that $E$$\approx$$E_{high-k}$$\approx$$E_{ISBT}$ \cite{agranovich_cavity_2003}. Taking this into account and solving for E to determine the resultant dispersion of the system from Equation \ref{AddingL0} we arrive at the following:
\begin{equation}
	\begin{split}
	E_{U,L}(q)=\frac{E_{high-k}(q)+E_{ISBT}}{2}\\
	\pm\frac{\sqrt{4(\Gamma)^2+(E_{high-k}(q)-E_{ISBT})^2}}{2}\
	\label{ClassicalSC}
	\end{split}
\end{equation}
Equation \ref{ClassicalSC} shows the formation of the resultant upper and lower polariton branches of the dispersion in the regime of strong coupling. The magnitude of the splitting between the upper and lower branches of the dispersion is proportional to $\Gamma^2 = \frac{C\hbar^2c^2k_{z}^2}{4\epsilon_{zz}\epsilon_{xx}E_{ISBT}E_{high-k}}$ and is much larger than the ISBT linewidth if the resonances are strongly coupled. The polaritonic dispersion emphasizes the mixing of the states between the two resonances in the strong coupling regime.

\subsection{Rabi Splitting in Semiconductor HMMs}
\label{StrongFormalism}
We now define the strong coupling behaviour in the semiconductor HMM through the Rabi splitting (RS) energy. The RS energy denotes the energy level splitting between two strongly coupled resonances within a system. The semiconductor HMM, as derived classically in Section \ref{BasisSC}, displays strong coupling phenomena when the energy of the ISBT and the high-$k$ mode become degenerate \cite{Dini2003}. The explicit regime of strong coupling occurs when the magnitude of the RS energy is greater than the sum of the linewidth of the high-$k$ mode and the radiative broadening of the ISBT resonance \cite{plumridge_ultra-strong_2007,skolnick_strong_1998,houdre_room-temperature_1994}. This results, as expected, in a mixed state between the two resonances of the system leading to a high-$k$-ISBT polariton.
\par The resultant dispersion and, more importantly, the magnitude of the splitting energy of the high-$k$-ISBT polariton can be accomplished by describing the coupling between two oscillators with a 2$\times$2 matrix Hamiltonian given by \cite{skolnick_strong_1998,houdre_room-temperature_1994}:
\begin{equation}
H=\left(\begin{array}{cc}
	E_{ISBT}&\frac{\hbar\Omega}{2}\\
	\frac{\hbar\Omega}{2}&E_{high-k}\end{array}\right)
	\label{Hamiltonian}
\end{equation}
Here, $E_{ISBT}$ and $E_{high-k}$ represent the respective energy dispersions of each of the resonances, specifically the ISBT and the high-$k$ mode respectively.  For the systems observed in this paper, the ISBT resonance is assumed to be at one particular energy across all values of the in-plane wavevector ($k_x$). The coupling matrix term proportional to $\hbar\Omega$ is representative of the Rabi splitting energy of the system. 

\par We solve the eigenvalue problem for the matrix given in Equation \ref{Hamiltonian} to determine the dispersion of the system \cite{skolnick_strong_1998,houdre_room-temperature_1994}: 

\begin{equation}
	\begin{split}
	E_{U,L}(q)=\frac{E_{high-k}(q)+E_{ISBT}}{2}\\
	\pm\frac{\sqrt{4(\frac{\hbar\Omega}{2})^2+(E_{high-k}(q)-E_{ISBT})^2}}{2}\
	\label{FormalSC}
	\end{split}
\end{equation}
Here we see solutions for the upper and lower branch of the polaritons observed from the strong coupling interaction between the two resonances. Comparing Equation \ref{ClassicalSC} and Equation \ref{FormalSC}, we clearly note that the RS energy ($\hbar\Omega$) has taken the place of the semi-classical splitting energy ($\Gamma$) in Equation \ref{ClassicalSC}. We can now define our splitting energy for the system with the known RS energy, $\hbar\Omega$, where $\Omega$ is the frequency corresponding to the RS. Equation \ref{FormalSC} assumes that $\hbar\Omega$, is much larger than the radiative broadening of the ISBT, as is the case in the strong coupling regime.

For the analysis done in this paper, we use Equation \ref{FormalSC} to determine the semiclassical Rabi splitting energy of the semiconductor HMM system. The semiclassical approach is warranted as the system does not deal with single emitters, but a multitude of emitters in the MQW layers. In Section \ref{SCEMT} and Section \ref{SCMULT} we will numerically determine the dispersion of the proposed semiconductor HMM as both an effective medium and a practical multilayer structure. We compare the analytical dispersion given by Equation \ref{FormalSC} to show that strong coupling is present in the system.

\section{Strong Coupling in Type II Semiconductor HMMs: Effective Medium Approach}
\label{SCEMT}
We now analyze the strong coupling interaction between the Type II high-$k$ modes of a semiconductor HMM and the ISBTs of the structure using effective medium theory. Metamaterials interacting with incident radiation at wavelengths much longer than the individual layer thicknesses of the structure can be homogenized and treated as an effective medium. 

\par The semiconductor HMM consists of a series of alternating subwavelength semiconductor layers (Figure \ref{Slab}(a)). Here, we show the transmission spectra of the MQW/In$_{0.53}$Ga$_{0.47}$As multilayer structure considered as an effective medium slab. We use the homogenized EMT equations for a uniaxial medium:

\begin{equation}
	\epsilon_{\parallel}=\epsilon_{InGaAs}\rho+(1-\rho)\epsilon_{xx}^d
	\label{EpsPara}
\end{equation}
\begin{equation}
	\frac{1}{\epsilon_\perp}=\frac{\rho}{\epsilon_{InGaAs}}+\frac{1-\rho}{\epsilon_{zz}^d}
	\label{EpsPerp}
\end{equation}
$\epsilon_{xx}^d$ and $\epsilon_{zz}^d$ are the parallel and perpendicular effective medium permittivities for the MQW slab respectively and $\epsilon_{InGaAs}$ is the permittivity of n$^+$-In$_{0.53}$Ga$_{0.47}$As. Note that the permittivities of the MQW slab ($\epsilon_{xx}^d$ and $\epsilon_{zz}^d$) are both positive while the In$_{0.53}$Ga$_{0.47}$As permittivity ($\epsilon_{InGaAs}$) is negative to achieve the hyperbolic dispersion of the slab. The fill fraction, $\rho$, is assumed to be 0.5 throughout the paper as both the MQW and n$^+$-In$_{0.53}$Ga$_{0.47}$As have equal layer thicknesses.

\begin{figure*}[ht]
	\includegraphics{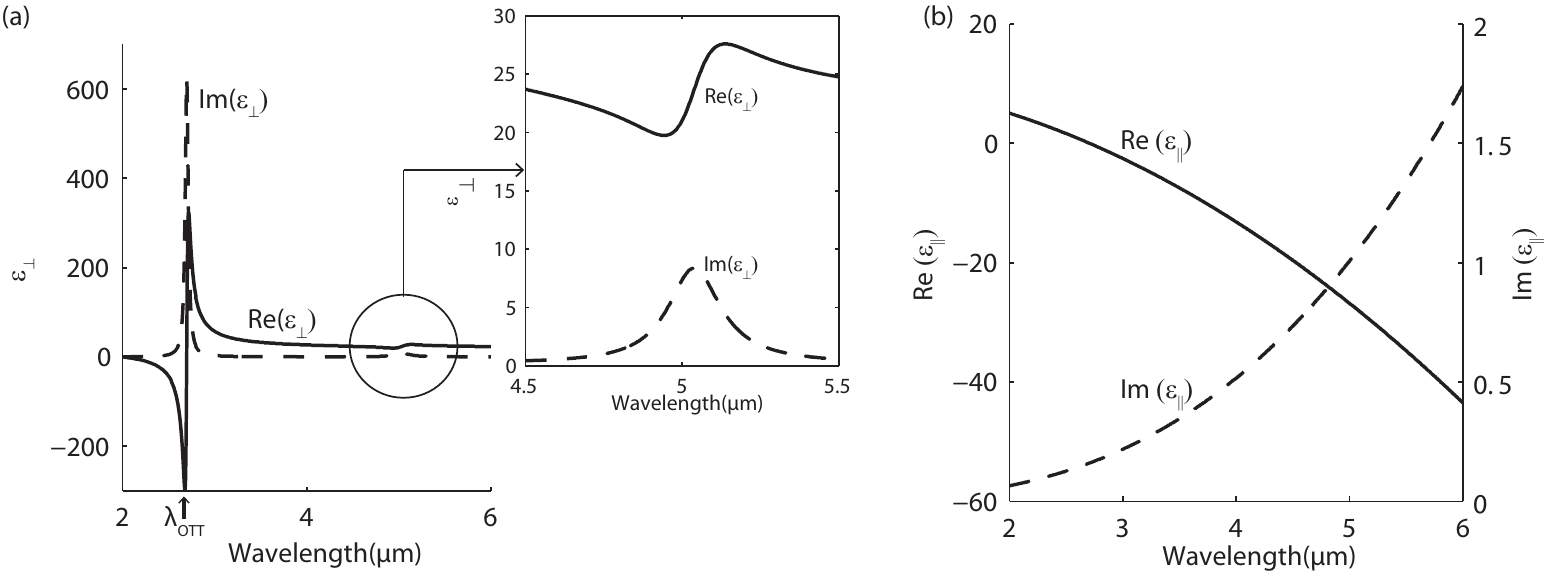}
	\caption{(a) Perpendicular and (b) parallel dispersions for the homogenized MQW-In$_{0.53}$Ga$_{0.47}$As structure given by Equation \ref{EpsPara} and Equation \ref{EpsPerp} respectively. A clear transition from the Type I to the Type II region is noticed at a wavelength of $\lambda_{OTT}\approx$ 2.8 $\mu$m where the parallel and perpendicular components of the permittivity both change sign. A smaller resonance at $\lambda_{ISBT}$ = 5 $\mu$m is shown in the inset of (a) which corresponds to the intersubband transition energy of the structure.}
	\label{HomoPerm}
\end{figure*}

\par The homogenized dispersions shown in Figure \ref{HomoPerm} are plotted for wavelengths larger than the plasma frequency and outline the transitions from the Type I region to the Type II region of the HMM. This shift in the dispersion of the metamaterial, where the two-sheeted hyperboloid (Type I) transitions to the single-sheeted hyperboloid (Type II), is a special case of an optical topological transition (OTT) \cite{krishnamoorthy_topological_2012}. In Figure \ref{HomoPerm}(a) we can see the resonance in the permittivity as result of the topological
transition at $\lambda_{OTT}\approx2.8$ $\mu$m.

\par Knowledge of $\lambda_{OTT}$ gives useful insight into the behaviour of our hyperbolic metamaterial for different regions of the electromagnetic spectrum. The semiconductor HMM is in the Type I region up to $\lambda_{OTT}\approx2.8$ $\mu$m after which point larger wavelengths correspond to a Type II HMM. Furthermore, the resonance in $\epsilon_\perp$ at $\lambda_{ISBT}$=5 $\mu$m (inset Figure \ref{HomoPerm}(a)) corresponds to the ISBT resonance of the structure in the Type II region. The imaginary component of the permittivity represents the material absorption. We can see in the inset of Figure \ref{HomoPerm}(a) that the imaginary permittivity is peaked at $\lambda_{ISBT}$=5 $\mu$m. Note that if the ISBT energy is tuned away from the range of wavelengths shown here, or is turned off completely, the resonance at $\lambda_{ISBT}$ would not appear in the dispersion of the perpendicular permittivity($\epsilon_\perp$). 

\begin{figure}[h]
	\includegraphics{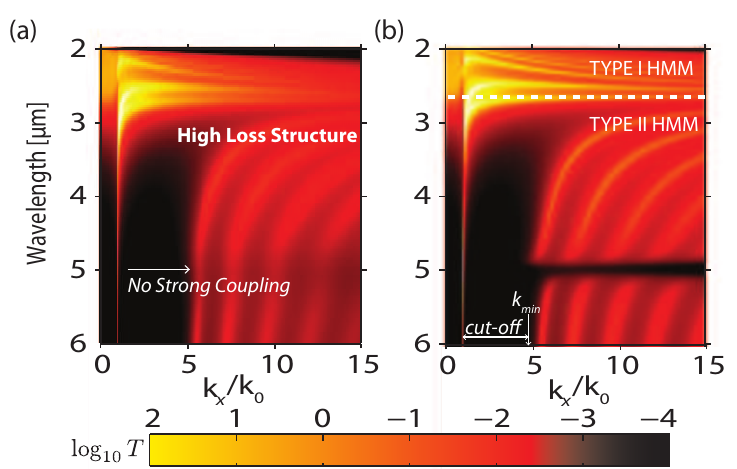}
	\caption{Transmission (log scale) through an 800 nm thick homogenized MQW-In$_{0.53}$Ga$_{0.47}$As slab simulated with the transfer matrix method. (a) Type II high-$k$ modes coupled to intersubband transitions (ISBTs) for a high loss semiconductor HMM slab. The electron scattering rate for both the MQW slab and the In$_{0.53}$Ga$_{0.47}$As layer was increased to $\gamma = 3.5\times10^{13}s^{-1}$ and $\gamma = 5\times10^{13}s^{-1}$ respectively to simulate a higher loss structure. No strong coupling takes place in this high loss regime. (b) Strong coupling between the ISBT and Type II modes of the slab at $\lambda_{ISBT}=5\mu$m for realistic scattering rates of the MQW slab ($\gamma = 7.6\times10^{12}s^{-1}$) and the In$_{0.53}$Ga$_{0.47}$As ($\gamma = 1\times10^{13}s^{-1}$) layer. A series of high-$k$-ISBT polaritons are formed. In both (a) and (b) the structure supports high-$k$ modes up to infinitely large wavevectors. The cut-off region indicates the wavevectors for which no transmission is allowed through the slab until $k_{min}$, indicating the smallest wavevector for  the 1st high-$k$ mode in the defined wavelength range. $k_0$ is the free-space wavevector.}
	\label{HomoTransmission}
\end{figure}

We use the dispersions shown in Figure \ref{HomoPerm} and the transfer matrix method to evaluate the transmission for an incident $p$-polarized plane wave through an 800 nm thick homogenized MQW-In$_{0.53}$Ga$_{0.47}$As slab surrounded by vacuum (Figure \ref{HomoTransmission} b). Figure \ref{HomoTransmission}(b) shows the series of bright bands that are the high-$k$ modes for the structure. We also note that due to the hyperbolic dispersion the in plane wavevector ($k_x$) is unbounded in this EMT limit, and high-$k$ modes up to wavevector magnitudes approaching infinity will be observed \cite{cortes_quantum_2012,Guo2012}. 

\par Closer examination of Figure \ref{HomoTransmission}(b) also shows distinguishable regions of the Type I and Type II modes in correspondence with the EMT parameters of Figure \ref{HomoPerm}. There is also a distinct cut-off region for the Type II modes where there is no transmission through the metamaterial in $k$-space. The metamaterial is highly metallic and thus extremely reflective in the cut-off region. The appearance of high-$k$ modes starts at the $k_{min}$ point where conditions are satisfied to support the high-$k$ modes for the structure \cite{cortes_quantum_2012}. 

\par We now turn our attention to the distinct feature of Figure \ref{HomoTransmission}(b) where each high-$k$ mode couples with ISBT resonance of the metamaterial showing anticrossing behaviour at the ISBT wavelength (($\lambda_{ISBT}= 5 \mu$m)). Subsequently, the high-$k$ mode gains a typical polariton like dispersion as a result of the strong coupling. The mixed state between the high-$k$ mode and the ISBT leads to the creation of an high-$k$-ISBT polariton. Strong coupling zones, whether it be particular wavevector regions or energies, can be assigned by tuning the ISBT energy or the dispersion profile of the modes \cite{Dini2003}. Both of these parameters can be tuned by the quantum well thickness and period as well as the doping density of the semiconductors in the structure. Note, however, that the dispersion of the permittivity and the losses of the systems would need to be taken into consideration in order to ensure that conditions for strong coupling are met. Figure \ref{HomoTransmission}(a) shows the transmission spectra for the semiconductor HMM where the material loss has been arbitrarily increased. No strong coupling between the high-$k$ modes and the ISBT takes place in this high loss regime. 

\par
\begin{figure}[h]
	\includegraphics{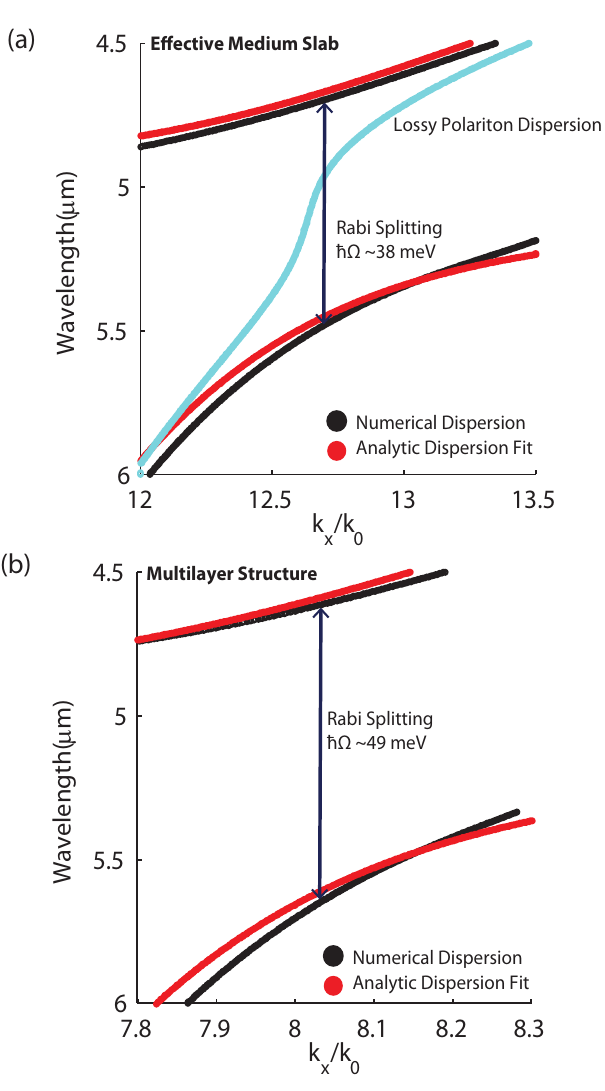}
	\caption{(a) Analytic (Equation \ref{FormalSC}) and numerical dispersions for the extracted 4$^{th}$ ‘high-$k$-ISBT polariton’ shown in Figure \ref{HomoTransmission}(b) for the EMT slab. The magnitude of the splitting energy is determined by using the RS energy as a fitting parameter in the analytical expression to match the numerical results. The fitting parameter  $\hbar\Omega$=38meV is used in the analytical expression. The 4$^{th}$ high-$k$ mode for the arbitrary high loss semiconductor HMM from Figure \ref{HomoTransmission}(a) is also shown and displays no strong coupling behaviour. (b) Extracted 3$^{rd}$ high-$k$-ISBT polariton for the multilayer semiconductor HMM shown in Figure \ref{TransmissionMult}(b). The fitting parameter  $\hbar\Omega$=49meV is used in the analytical expression.}
	\label{SimulationSplit}
\end{figure}

 The magnitude of splitting observed can be quantified by extracting a specific high-$k$-ISBT polariton from the dispersions in Figure \ref{HomoTransmission}(b) and matching it to the analytical expression of Equation \ref{FormalSC}. The RS energy in Equation \ref{FormalSC} can be used as a fitting parameter to achieve the best fit between the numerical results and the analytical expression. The 4$^{th}$ high-$k$ ISBT polariton (for the region lying between 12-13.5 $k_x/k_0$)  is extracted, as seen in Figure \ref{SimulationSplit}(a), and plotted in conjunction with the analytical expression with a fitting parameter for the RS energy at $\hbar\Omega$ = 38meV.   
\par Figure \ref{SimulationSplit} provides a visual of the numerical simulation and analytic model for the strong coupling in the semiconductor HMM. There is a strong correlation between the numerical transfer matrix dispersion and the 2 level model analytic dispersion (Equation \ref{FormalSC}) using the RS energy as a fitting parameter. This allows us to make a good approximation of the splitting energy for the 4$^{th}$ high-$k$-ISBT polariton. In addition to the strongly coupled ISBT and high-$k$ mode, we show the 4$^{th}$ high-$k$ mode for the high loss semiconductor HMM from Figure \ref{HomoTransmission}(a). We can clearly see that no strong coupling takes place in the high loss regime and that the lower polariton branch back-bends toward the top branch. 

\begin{table}
	\centering
	\caption{Rabi splitting (RS) energy between Type II HMM Modes and the ISBT for the 800 nm thick homogenized MQW-In$_{0.53}$Ga$_{0.47}$As slab shown in Figure \ref{HomoTransmission}. The magnitude of the RS is decreasing with increasing Type II mode number and wavevector magnitude ($k_x/k_0$ ).} 
	\begin{tabular}{c|c|c}
	\hline\hline
	High-$k$ ISBT & $k_x/k_0$ Bounds & RS Energy ($\hbar\Omega$) [meV]\\ \hline
	1 & 5-6.6 & 45 \\
	2 & 6.9-8.7 & 41\\
	3 & 9-11.2 & 39 \\
	4 & 12-15 & 38\\
	\hline\hline
	\end{tabular}
	\label{Table1}
\end{table} 

\par The approximated RS energies for all the high-$k$-ISBT polaritons (Table \ref{Table1}) show that the maximum splitting occurs for the first polariton, with a RS energy approximately 9 times greater than the ISBT linewidth. The magnitude of the RS decreases with increasing wavevector magnitude due to increased confinement of the high-$k$ modes and therefore less mode overlap with the MQW structure. This is sufficient to satisfy the strong coupling requirement between the high-$k$ states and the ISBT. 

\par It is important to realize that if the total losses in the system were greater than the degree of interaction between the high-$k$ mode and the ISBT (as determined by the RS energy) no strong coupling would take place. In the semiconductor HMM presented here, the energy losses corresponding to electron scattering and radiative broadening of the ISBT are 6.6 meV and 5 meV respectively. We see in Table \ref{Table1} that the smallest magnitude of the RS energy (38 meV) is sufficiently larger than the total energy loss in the system (11.6 meV). As a result, each of the high-$k$ modes strong couples to the ISBT.  

\section{Strong Coupling in Type II Semiconductor HMMs: Multilayer Realization}
\label{SCMULT}
We now validate the EMT calculations of Section \ref{SCEMT} with a practical multilayer approach for the semiconductor HMM. Here, we determine the transmission of the incident radiation through each individual layer of the structure with the transfer matrix method. Determination of optical properties in this fashion is more representative of a structure conceived in fabrication.

\begin{figure}[h]
	\includegraphics{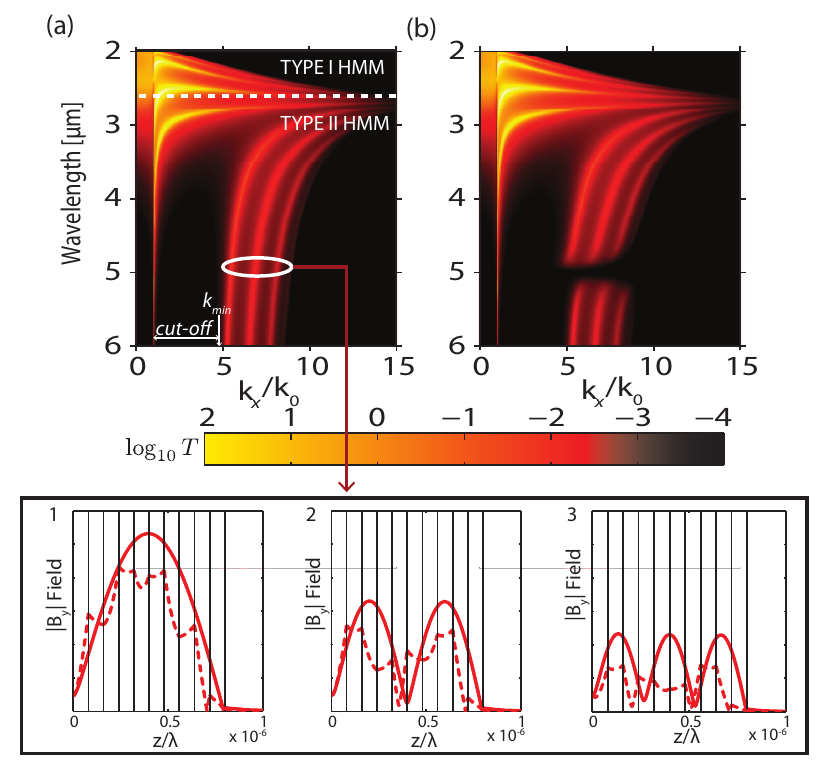}
	\caption{Transmission of 10 alternating MQW and In$_{0.53}$Ga$_{0.47}$As 80 nm layers using the transfer matrix method. (a) Type I and Type II high-$k$ modes of the multilayer structure with intersubband transitions (ISBTs) tuned away from the mode energies. (b) Strong coupling between the ISBT and Type II modes of the multilayer structure at $\lambda_{ISBT}=5\mu$m. A series of high-$k$-ISBT polaritons are formed.  In (b) the multilayer structure shows agreement with the effective medium results of Figure \ref{HomoTransmission}(b). Note that the multilayer structure, in comparison to the homogenized slab, experiences an upper cutoff of the wavevector as it approaches the size of the unit cell. The cut-off region for wavevectors smaller than $k_{min}$ shows a match to EMT (Figure \ref{HomoTransmission}). $k_0$ is the free-space wavevector. Inset of (a) shows relative magnitudes of the in-plane magnetic field ($|B_y|$) at a wavelength of 5 $\mu$m for the first 3 high-$k$ modes of the  MQW-In$_{0.53}$Ga$_{0.47}$As multilayer(dashed) and homogenized(solid) structure.}
	\label{TransmissionMult}
\end{figure}

Strong coupling behaviour in a practical multilayer realization of the semiconductor HMM shows comparable results to those seen with EMT. Analysis of the transmission spectra of the semiconductor HMM was obtained through the numerical transfer matrix method (Figure \ref{TransmissionMult}). The multilayer structure analyzed consists of 5 layers of an 80 nm MQW slab alternated with 5 layers of an 80 nm n$^+$-In$_{0.53}$Ga$_{0.47}$As semiconductor for a total structure thickness of 800 nm. Note that the total thickness of the structure is the same thickness as the analysis done with the EMT slab in Section \ref{SCEMT}. 

\par Figure \ref{TransmissionMult} shows agreement with the EMT dispersions shown in Figure \ref{HomoTransmission}, including the mode profile of the high-$k$ modes as well as the strong coupling behaviour with the ISBT. There do exist some limitations with EMT that can account for the differences between Figure \ref{HomoTransmission} and Figure \ref{TransmissionMult} \cite{kidwai_effective-medium_2012,guclu_hyperbolic_2012}. For example, the multilayer structure has a distinct upper cut-off for the high-$k$ modes that was not seen in the EMT structure. At larger wavevector magnitudes the waves begin propagating with wavelengths comparable to the size of the unit cell and no longer interact with the structure as an effective medium. The wavevectors lie at the edge of the brillioun zone of the periodic lattice and begin to bragg scatter, leading to an upper limit to the wavevector magnitude of the high-$k$ modes which can propagate in the multilayer structure. The transfer matrix method takes the size of the unit cell into account, and thus in the multilayer semiconductor HMM (Figure \ref{TransmissionMult}), a sharp upper cut-off is observed at the point where the wavevector becomes comparable to the unit cell size \cite{cortes_quantum_2012}.

\par Figure \ref{TransmissionMult}(b) confirms the strong coupling behaviour in the multilayer semiconductor HMM. The magnitude of the strong coupling in the multilayer structure was extracted by comparing the analytical expression of Equation \ref{FormalSC} with the numerical results in the same manner as was done with the EMT slab (Figure \ref{SimulationSplit}(b)). We note that in the multilayer structure only 3 high-$k$-ISBT polaritons are present in comparison to the 4 seen with the EMT slab of the same thickness due to the upper cut-off wavevector of the multilayer structure. As expected, the maximum RS energy ($\hbar\Omega$= 52 meV) occurs for the first high-$k$-ISBT polariton for the system, a value approximately 10.5 times greater than the ISBT linewidth and 4.5 times greater than the material loss of the structure. The Rabi splitting is larger in the multilayer structure as a result of local field enhancements of the discontinuous $E_z$ fields at each interface. This is not observed in the EMT slab.

\begin{table}
	\centering
	\caption{Rabi splitting (RS) energy between Type II HMM Modes and the ISBT for the MQW-In$_{0.53}$Ga$_{0.47}$As multilayer structure shown in Figure \ref{TransmissionMult}. The magnitude of the RS is decreasing with increasing Type II mode number and wavevector, similar to the results  seen in Table \ref{Table1} for the homogenized MQW-In$_{0.53}$Ga$_{0.47}$As slab}
	\begin{tabular}{c|c|c}
	\hline\hline
	High-$k$ ISBT & $k_x/k_0$ Bounds & RS Energy ($\hbar\Omega$) [meV]\\ \hline
	1 & 4.5-6.2 & 52\\
	2 & 6.6-7.6 & 50\\
	3 & 7.8-8.8 & 49\\
	\hline\hline
	\end{tabular}
	\label{Table2}
\end{table} 

\par Upon closer inspection of Table \ref{Table1} and \ref{Table2}, we see the magnitude of the RS energy decreases with increasing values of the in-plane wavevector ($k_x$) for both the EMT slab and the multilayer structure. This is explained by observing that the maximum amplitude of the electric fields in the growth direction ($E_z$) also decreases with $k_x$ and the Type II mode number. This is outlined in the inset of Figure \ref{TransmissionMult}(a) using the in-plane magnetic fields ($|B_y|$). We use ($|B_y|$) instead of the discontinuous perpendicular electric fields ($|E_z|$) for the sake of clarity. The ISBT, for the coordinate axis used in this paper, requires z-polarized E-fields for the transition to be allowed as a result of orthogonality conditions. The strength of the ISBT is dependent on the magnitude of the electric fields normal to the interface and, as a result, the decreasing $E_z$ field magnitude leads to a decreased ISBT absorption. The decreased strength of the transition leads to reduced coupling with the high-$k$ mode and the overall RS energy is decreased.

In order to study the predicted strong coupling effect experimentally it will be necessary to probe the high $k$ ($k_x/k_0>$3) regions of the structure. The fact that we are operating in the 1-5 $\mu$m wavelength regime allows for the use of higher index materials such as silicon (n$\approx$3.5) that can be used to prism couple into the high-$k$ states. Additionally, the refractive index of the dielectric layers in the structure itself are also relatively high (n$\approx$3-3.5) which provides a large degree of tunability. The high-$k$ modes of the HMM can also be shifted to lower $k_x/k_0$ regions by reducing the metallic character of the semiconductors with decreased doping. Grating coupling methods are also a viable option to couple to much larger values of $k_x/k_0$ in the structure.

\section{Conclusion}

In this paper, we have described strong coupling interactions between the high-$k$ modes of the HMM and the intersubband transitions of the embedded quantum wells with Rabi splitting energies up to 52 meV (approximately 10.5 times greater than the ISBT linewidth). The system showed strong coupling behaviour in the effective medium approach as well as a practical structure. This is the first example of strong coupling behaviour in hyperbolic metamaterials. Prism coupling is necessary to couple incident light into the high-$k$ modes of the metamaterial and experimentally verify our predicted effect. This structure can have potential applications in quantum well infrared photodetectors and tunable intersubband light-emitting devices.

\appendix

\section*{Appendix A: Semi-classical Strong Coupling}

Here, we derive in more detail the form of Equation \ref{ClassicalSC}. Starting from Equation \ref{AddingL0} and subbing in our dispersion for our high-$k$ mode energy given by Equation \ref{high-kEnergy}, we can rearrange our new expression in the following manner as shown by Equation \ref{A1}.

\begin{equation}
	\frac{E^2_{high-k}-\frac{\hbar^2c^2k_{z}^2}{\epsilon_{xx}}}{E^2-\frac{\hbar^2c^2k_{z}^2}{\epsilon_{xx}}} - \frac{C}{\epsilon_{zz}(E_{ISBT}^2-E^2)}=1
	\tag{A1}
	\label{A1}
\end{equation}

If we now let $\alpha^2=\frac{\hbar^2c^2k_{z}^2}{\epsilon_{xx}}$, we can now express Equation \ref{A1} in the form of Equation \ref{A2}:

\begin{equation}
	\begin{split}
	\epsilon_{zz}(E_{high-k}-\alpha)(E_{high-k}+\alpha)(E_{ISBT}-E)(E_{ISBT}+E)\\
	- C(E-\alpha)(E+\alpha) = \epsilon_{zz}(E^2-\alpha^2)(E_{ISBT}^2-E^2)\
	\end{split}
	\label{A2}
	\tag{A2}
\end{equation}

As outlined in Section \ref{BasisSC} we know $E$$\approx$$E_{high-k}$$\approx$$E_{ISBT}$ in the strong coupling regime. As such, we can substitute the following expressions into our equations: $E_{ISBT}+E$$\approx$$2E$ and $E_{high-k}+E$$\approx$$2E$. Further algebra then leads to the expression given by Equation \ref{A3}:

\begin{equation}
	\tag{A3}
	\begin{split}
		4E^2(E_{ISBT}E_{high-k}-E_{ISBT}E-E_{high-k}E+E^2)\\
		=\frac{C}{\epsilon_{zz}}(E^2-\alpha^2)\
	\label{A3}
	\end{split}
\end{equation}

If we now substitute in $\alpha$ into Equation \ref{A3} and put it in quadratic form, we arrive at the equation below:

\begin{equation}
	\tag{A4}
	\label{A4}
	\begin{split}
	E^2-E(E_{ISBT}+E_{high-k}) \\
	+ E_{ISBT}E_{high-k}=\frac{C\hbar^2c^2k_{z}^2}{4\epsilon_{zz}\epsilon_{xx}E_{ISBT}E_{high-k}}\
	\end{split}
\end{equation}

If we now let $\Gamma^2 = \frac{C\hbar^2c^2k_{z}^2}{4\epsilon_{zz}\epsilon_{xx}E_{ISBT}E_{high-k}}$ and solve the resultant quadratic equation for E, we arrive at the expression for our upper and lower polariton branches given by Equation \ref{ClassicalSC}. Note the  system will be inevitably be curtailed by loss thus negating any singularities in the above expressions.

\section*{Appendix B: Rabi Splitting Dispersion}

Here we show how the form of Equation \ref{FormalSC} is obtained from the matrix Hamiltonian given in Equation \ref{Hamiltonian} for the coupling between $E_{high-k}$ and $E_{ISBT}$. The expression for the energy dispersion of the upper and lower polariton branches is done by simply finding the eigenvalues of Equation \ref{Hamiltonian} by setting the determinant of $H-EI$ to 0, where $I$ is the identity matrix:

\begin{equation}
	\tag{B1}
	\label{B1}
	0=det\left[\begin{array}{cc}
	E_{ISBT}-E&\frac{\hbar\Omega}{2}\\
	\frac{\hbar\Omega}{2}&E_{high-k}-E\end{array}\right]
\end{equation}
\begin{equation}
	\tag{B2}
	\label{B2}
	\begin{split}
	E^2-E(E_{high-k}+E_{ISBT}) \\
	- (\frac{\hbar\omega}{2})^2+E_{ISBT}E_{high-k}=0	
	\end{split}
\end{equation}

By setting the determinant in Equation \ref{B1} to 0, we get the resultant quadratic equation shown in Equation \ref{B2}. Solving and simplying for $E$ in Equation \ref{B2} results in the expresion shown in Equation \ref{FormalSC} giving us our expression for the energy dispersions of the upper and lower polariton branches.

\bibliographystyle{unsrt}
\bibliography{Strong_Coupling_Ref4}

\end{document}